# A Thermoelastoplastic Material Model for Finite-Strain Cyclic Plasticity of Metals

Ladislav Écsi[1], Péter Ván[2,4,5,*], Tamás Fülöp[2,5], Balázs Fekete[3], Pavel Élesztős[1], Roland Jančo[1]

[1] STU in Bratislava, Faculty of Mech. Eng., Námestie slobody 17, 812 31 Bratislava 1, SK

[2] BME, Dept. of Energy Engineering, Faculty of Mechanical Engineering, , Bertalan Lajos utca 4-6, H-1111 Budapest, HU

[3] University of Dunaújváros, Táncsics Mihály utca 1/a, H-2401 Dunaújváros, HU

[4] MTA Wigner RCP, Dep. of Theoretical Physics, Konkoly Thege Miklós út 29-33, 1121 Budapest, HU

[5] Montavid Thermodynamic Research Group, Igmándi út 26, H-1112 Budapest, HU

[*] van.peter@wigner.mta.hu

**Abstract:** This paper presents an improved version of our former material model capable of modelling ductile-to brittle failure mode transition in ductile material undergoing deformations at high strain rates, and demonstrates the model in a numerical study using a fully coupled thermal-structural finite element analysis of a notched aluminium alloy specimen loaded in cyclic tension. The model is based on an objective representation of the deformation and stress measures and based on rate type constitutive equations. It does not only complies with the principles of material modelling, but it also uses constitutive equations, evolution equations and even 'normality rules' during return mapping which can be expressed in terms of power conjugate stress and strain measures, or their objective rates.

## 1. Introduction

The modelling of plastic behaviour of structural materials in the framework of finite-strain flow plasticity theories do not perform well when compared with experiments [1]. Although to date there have been proposed a lot of material models for finite-strain elastoplasticity (see e.g. [2-10], the models in general lack universality, as their analysis results depend on the description used in the model and the particularities of the model formulation. It may well be that the modelling methodology simply needs some developments in order for the related theories to be considered complete. The consideration of thermal phenomena is exceptionally challenging when internal variables are to be considered.

Contemporary flow plasticity theories in finite-strain, phenomenological thermoelastoplasticity are either based on the additive decomposition of a strain rate tensor into an elastic part, a plastic part, or the multiplicative decomposition of a deformation gradient into an elastic part and a plastic part. Thermal phenomena are usually introduced by thermoelastic equations of state. . The Theories with additive strain rate decomposition are considered to be ad hoc extensions of infinitesimal flow plasticity theories into the area of finite deformations of elastic media to cover large displacements, but small strains of the deforming body. The thermoelastic behaviour is traditionally introduced throught a particular thermodynamics potentials, representing the thermoelastic equation of state of the material [33,34]. The related material models use also the additive decomposition of the strain rate tensor into an elastic part, a plastic part and are based on a hypoelastic stress-strain relationship while utilizing the nonlinear continuum-mechanical theory of elastic media to describe the plastic behaviour of the material [2, 11-16]. Without internal variables the thermal and elastoplastic parts are not directly coupled in isotropic material [35].

The second type of flow plasticity theories is now generally accepted as "proper finite-strain flow plasticity theories", utilizing the theory of single-crystal plasticity to describe the micromechanics of irreversible deformations in the material. The related material models are based on the multiplicative split of a deformation gradient into an elastic part, a plastic part and a thermal part, and employ classic flow plasticity models from small-strain elastoplasticity while utilizing the nonlinear continuum-mechanical theory of finite deformations of elastic media to describe the plastic flow in the material [2-4, 17-23] The treatment of thermal phenomena is similar to the case of additive decomposition: thermodynamic potentials play the main role.

Our ongoing research, however, has shown that in a reference frame independent theoretical framework the additive decomposition is the proper kinematic approach also in case of finite deformations (Fülöp and Ván 2012). Moreover, further investigations has shown that thermoleasticity is conveniently and reasonably represented introducing the thermal strain concept. In a reference frame independent treatment thermal strain rate is additively decomposed to elastic and plastic strain rates [36]. An other important aspect of the theory is that it does not assume the hypothetical stress free state of the continuum. Our ongoing research introduced an easier introduced a lightweight and transparent treatment, where an additive decomposition of the displacement field into an elastic part, a plastic part and a thermal part, and which describes plastic flow in terms of various instances of a yield surface and stress measures in the initial and current configurations of the body. The theory moreover allows for the generalization of contemporary flow plasticity theories and the development of thermodynamically consistent material models.

The aim of this paper is to present an alternative $J_2$ material model for finite-strain cyclic plasticity of metals using combined kinematic and isotropic hardening, which in addition can predict ductile-to-brittle failure mode transition in the material undergoing deformations at high strain rates and predict the corresponding thermal behaviour. We will see, that thermal strain is best described by a spherical tensor representation, expressing the volumetric nature of thermoelastic phenomena. The model is based on the first nonlinear continuum theory of finite deformations of elastoplastic media, which allows for a thermodynamically consistent description of the plastic flow. Though the continuum theory is not detailed herein, it will be shown that contemporary strain rate tensor additive decomposition-based flow plasticity theories are in fact finite-strain theories, but they are constrained when the plastic flow in them is defined in terms of a Cauchy stress tensor based yield surface in the current configuration of the body. In this approach the numerical treatment of the thermal behaviour is simple and results in stable and fast convergent numerical schemes

**2. Theory**

The Lagrangian description is used to describe the motion of a material particle of a deforming body. Though a single form using a particular stress measure is sufficient to formulate a thermodynamically consistent material model, all forms of the model, using various stress measures, are needed in order to prove that the formulation of the model is thermodynamically consistent.

**2.1 Kinematics**

Starting with a multiplicative split of the deformation gradient, the proper form of the decomposition, consistent with the theory of nonlinear continuum mechanics, which includes displacement fields in the definitions of the gradient and its parts, can be expressed as (see Figure 1)

**Fig. 1.** *Multiplicative decomposition of the deformation gradient into an elastic part, a plastic part and a thermal part*

$$\mathbf{F} = \nabla_0(\mathbf{x}) = \frac{\partial \mathbf{x}}{\partial \mathbf{X}} = \frac{\partial \mathbf{x}}{\partial^{i2}\mathbf{X}} \cdot \frac{\partial^{i2}\mathbf{X}}{\partial^{i1}\mathbf{X}} \cdot \frac{\partial^{i1}\mathbf{X}}{\partial \mathbf{X}} = \left(\mathbf{I} + \frac{\partial^{i2}\mathbf{u}^{el}}{\partial^{i2}\mathbf{X}}\right) \cdot \left(\mathbf{I} + \frac{\partial^{i1}\mathbf{u}^{pl}}{\partial^{i1}\mathbf{X}}\right) \cdot \left(\mathbf{I} + \frac{\partial^{0}\mathbf{u}^{th}}{\partial \mathbf{X}}\right) =$$
$$= \mathbf{F}^{el}\left(^{i2}\mathbf{X}, t\right) \cdot \mathbf{F}^{pl}\left(^{i1}\mathbf{X}, t\right) \cdot \mathbf{F}^{th}\left(\mathbf{X}, t\right). \quad (1)$$

In this form, however, the deformation gradient has neither a Lagrangian form nor an Eulerian form, because the thermal displacement field $^0\mathbf{u}^{th} = {^0\mathbf{u}^{th}}(\mathbf{X},t)$, defined over the initial volume $^0\Omega$ of the body, is in Lagrangian form, but the elastic $^{i2}\mathbf{u}^{el} = {^{i2}\mathbf{u}^{el}}(^{i2}\mathbf{X},t)$ and plastic $^{i1}\mathbf{u}^{pl} = {^{i1}\mathbf{u}^{pl}}(^{i1}\mathbf{X},t)$ displacement fields, defined over the intermediate volumes $^{i2}\Omega$, $^{i1}\Omega$ of the body, are in Eulerian forms. In order that the multiplicative split be in Lagrangian form, the elastic and plastic displacement fields must have Lagrangian counterparts, $^{i2}\mathbf{u}^{el} = {^0\mathbf{u}^{el}} = {^0\mathbf{u}^{el}}(\mathbf{X},t)$, $^{i1}\mathbf{u}^{pl} = {^0\mathbf{u}^{pl}} = {^0\mathbf{u}^{pl}}(\mathbf{X},t)$, defined over the initial volume of the body, so that the proper Lagrangian form of the multiplicative split of the deformation gradient is actually based on the additive decomposition of the Lagrangian displacement field into an elastic part, a plastic part and a thermal part $^0\mathbf{u} = {^0\mathbf{u}^{th}} + {^0\mathbf{u}^{pl}} + {^0\mathbf{u}^{el}}$, where

$$^0\mathbf{u}^{th} = \left[\int_{X_{ref}}^{X} \alpha_X \cdot \left[{^0T}(\mathbf{X},t) - T_{ref}\right] \cdot dX, \int_{Y_{ref}}^{Y} \alpha_Y \cdot \left[{^0T}(\mathbf{X},t) - T_{ref}\right] \cdot dY, \int_{Z_{ref}}^{Z} \alpha_Z \cdot \left[{^0T}(\mathbf{X},t) - T_{ref}\right] \cdot dZ\right]^T. \quad (2)$$

In Eqn. (2) $\alpha_X, \alpha_Y, \alpha_Z$ denote the coefficients of thermal expansion defined in the initial configuration of the body and $^0T = {^0T}(\mathbf{X},t), T_{ref}$ are the material temperature field and the reference temperature respectively.

It ought to be mentioned here that the critics of the theory using the multiplicative split of the deformation gradient alone may have argued about the correctness of the theory from the physical point of view, since the multiplicative split predefines the order of deformations. According to Eqn. (1) the body firstly undergoes thermal deformations, secondly plastic deformations and lastly elastic deformations at each of its constituents. Obviously a virgin material will never deform in this way, since no plastic deformations can develop in it without prior elastic deformations. The inclusion of the displacement fields into the definition of the deformation gradient and their parts then puts the theory in order even from the physical point of view, since vector addition is commutative, resulting in the final Lagrangian form for the deformation gradient Eqn.(3), independent from the order of elastic, plastic and thermal deformations.

$$\mathbf{F}(\mathbf{X},t) = \mathbf{F}^{el}(\mathbf{X},t) \cdot \mathbf{F}^{pl}(\mathbf{X},t) \cdot \mathbf{F}^{th}(\mathbf{X},t) = \left[ \mathbf{I} + \frac{\partial^0 \mathbf{u}^{el}}{\partial \mathbf{X}} \cdot \left( \mathbf{I} + \frac{\partial^0 \mathbf{u}^{th}}{\partial \mathbf{X}} + \frac{\partial^0 \mathbf{u}^{pl}}{\partial \mathbf{X}} \right)^{-1} \right] \cdot$$
$$\cdot \left[ \mathbf{I} + \frac{\partial^0 \mathbf{u}^{pl}}{\partial \mathbf{X}} \cdot \left( \mathbf{I} + \frac{\partial^0 \mathbf{u}^{th}}{\partial \mathbf{X}} \right)^{-1} \right] \cdot \left( \mathbf{I} + \frac{\partial^0 \mathbf{u}^{th}}{\partial \mathbf{X}} \right) = \mathbf{I} + \frac{\partial^0 \mathbf{u}^{th}}{\partial \mathbf{X}} + \frac{\partial^0 \mathbf{u}^{pl}}{\partial \mathbf{X}} + \frac{\partial^0 \mathbf{u}^{el}}{\partial \mathbf{X}}.$$

(3)

In this case, however, neither the Green strain tensor $\mathbf{E} = 1/2 \cdot (\mathbf{F}^T \cdot \mathbf{F} - \mathbf{I})$, nor the Almansi strain tensor $\mathbf{e} = 1/2 \cdot (\mathbf{I} - \mathbf{F}^{-T} \cdot \mathbf{F}^{-1})$ has decomposition into an elastic part, a plastic part and a thermal part, but additive decomposition exists when one evaluates the objective rates of the strain tensors. As a result, the material $\dot{\mathbf{E}}$ and the spatial $\mathbf{d} = \mathcal{L}_e(\mathbf{e})$ strain rate tensors can be expressed in the following forms:

$$\dot{\mathbf{E}} = \frac{1}{2} \cdot (\dot{\mathbf{F}}^T \cdot \mathbf{F} + \mathbf{F}^T \cdot \dot{\mathbf{F}}) = \dot{\mathbf{E}}^{el} + \dot{\mathbf{E}}^{pl} + \dot{\mathbf{E}}^{th},$$ (4)

where

$$\dot{\mathbf{E}}^{el} = \frac{1}{2} \cdot \left[ \left( \frac{\partial^0 \dot{\mathbf{u}}^{el}}{\partial \mathbf{X}} \right)^T \cdot \mathbf{F} + \mathbf{F}^T \cdot \frac{\partial^0 \dot{\mathbf{u}}^{el}}{\partial \mathbf{X}} \right],$$ (5)

$$\dot{\mathbf{E}}^{pl} = \frac{\dot{\lambda}}{2} \cdot \left[ \left( \frac{\partial^P \Psi}{\partial \mathbf{P}} \right)^T \cdot \mathbf{F} + \mathbf{F}^T \cdot \frac{\partial^P \Psi}{\partial \mathbf{P}} \right],$$ (6)

$$\dot{\mathbf{E}}^{th} = \frac{1}{2} \left[ \begin{pmatrix} \alpha_X & 0 & 0 \\ 0 & \alpha_Y & 0 \\ 0 & 0 & \alpha_Z \end{pmatrix} \cdot \mathbf{F} + \mathbf{F}^T \cdot \begin{pmatrix} \alpha_X & 0 & 0 \\ 0 & \alpha_Y & 0 \\ 0 & 0 & \alpha_Z \end{pmatrix} \right] \cdot {}^0\dot{T}, \quad \frac{\partial^0 \dot{\mathbf{u}}^{th}}{\partial \mathbf{X}} = \begin{pmatrix} \alpha_X & 0 & 0 \\ 0 & \alpha_Y & 0 \\ 0 & 0 & \alpha_Z \end{pmatrix} \cdot {}^0\dot{T},$$ (7)

$$\frac{\partial^0 \dot{\mathbf{u}}^{pl}}{\partial \mathbf{X}} = \dot{\lambda} \cdot \frac{\partial^P \Psi}{\partial \mathbf{P}} = -\dot{\lambda} \cdot \frac{\partial^P \Psi}{\partial^P \mathbf{X}}, \quad \text{and} \quad \frac{\partial^P \Psi}{\partial \mathbf{P}} \neq \left( \frac{\partial^P \Psi}{\partial \mathbf{P}} \right)^T,$$ (8)

$$\mathbf{d} = \mathbf{d}^{el} + \mathbf{d}^{pl} + \mathbf{d}^{th}, \quad \mathbf{d}^{el} = \mathbf{F}^{-T} \cdot \dot{\mathbf{E}}^{el} \cdot \mathbf{F}^{-1}, \quad \mathbf{d}^{pl} = \mathbf{F}^{-T} \cdot \dot{\mathbf{E}}^{pl} \cdot \mathbf{F}^{-1}, \quad \mathbf{d}^{th} = \mathbf{F}^{-T} \cdot \dot{\mathbf{E}}^{th} \cdot \mathbf{F}^{-1}.$$ (9)

Here $\mathbf{X}$ denotes the position vector of a material point and $\mathbf{x} = \mathbf{X} + {}^0\mathbf{u}$ is the position vector of the corresponding spatial point after deformation. Then the deformation gradient $\mathbf{F} = \mathbf{I} + \partial^0 \mathbf{u}/\partial \mathbf{X} =$

$= \mathbf{I} + \partial\, ^0\mathbf{u}^{el}/\partial \mathbf{X} + \partial\, ^0\mathbf{u}^{pl}/\partial \mathbf{X} + \partial\, ^0\mathbf{u}^{th}/\partial \mathbf{X}$ can be expressed either as a function of the material displacement field $^0\mathbf{u}$ alone, or as a function of its elastic $^0\mathbf{u}^{el}$, plastic $^0\mathbf{u}^{pl}$ and thermal $^0\mathbf{u}^{th}$ parts. The symbols $\dot{\mathbf{E}}^{el}, \dot{\mathbf{E}}^{pl}, \dot{\mathbf{E}}^{th}/\mathbf{d}^{el}, \mathbf{d}^{pl}, \mathbf{d}^{th}$ denote the elastic, the plastic and the thermal material/spatial strain rate tensors, where in the second the plastic flow is defined by Eqn. (8)$_1$ as a product of a plastic multiplier $\dot{\lambda}$ and an appropriate yield surface normal $\partial\, ^P\Psi/\partial \mathbf{P}$, or $\partial\, ^P\Psi/\partial\, ^P\mathbf{X}$, either in terms of the 1$^{st}$ Piola-Kirchhoff stress tensor $\mathbf{P}$, or the corresponding backstress tensor $^P\mathbf{X}$, which is also a 1$^{st}$ Piola-Kirchhoff stress measure. It ought to be noted too that the calculation of the material thermal strain rate tensor (Eqn. (7)$_1$), in which there were neglected the off-diagonal elements of the material gradient of the thermal velocity field (Eqn. (7)$_2$), has been simplified since the contribution of the off-diagonal elements is small, unless the material gradient of the rate of the temperature field is high within the element. Here the symbol $\mathcal{L}_e(\mathbf{e})$ denotes the Lie derivative of the Almasi strain tensor $\mathbf{e}$, defined in terms of the Lie derivative operator of a spatial strain tensor $\mathcal{L}_e(\bullet) = \mathbf{F}^{-T} \cdot \left[\partial\left(\mathbf{F}^T \cdot (\bullet) \cdot \mathbf{F}\right)/\partial t\right] \cdot \mathbf{F}^{-1}$. It should also be noted that all the elastic, plastic and thermal strain rate tensors have forms similar to the strain rate tensor itself. Moreover, the plastic flow defined by Eqn. (8)$_1$ is not constrained, resulting in Eqns. (6) and (9)$_3$ respectively being the only non-degenerated forms of the material and spatial plastic strain rate tensors.

## 2.2 The constitutive and evolution equations of the material

The proper formulation of a material model for finite-strain elastoplasticity allows for the definition of the constitutive and evolution equations of the material in terms of various stress and strain measures or their objective rates. As a result, the equations cannot have unique forms. Moreover, all forms have to comply with the principles of material modelling, particularly to meet the requirements of material objectivity and be thermodynamically consistent, so that the equations can define the same material [24]. Furthermore, because the additive decomposition defined by Eqns. (4), (9)$_1$ exists in rate forms only, the constitutive and evolution equations must have rate forms too. In fact, Eqns. (8)-(11) define a true hypoelastic-based elastoplastic material model, which does not have a constitutive equation in terms of a finite strain measure.

In this study our former material model has been modified to be able to imitate ductile-to-brittle failure mode transition in a ductile material undergoing deformations at high strain rates [25]. With respect to the above, the rate form of the constitutive equation of the material can be expressed in any of the following forms in the 2$^{nd}$ Piola-Kirchhoff, 1$^{st}$ Piola-Kirchhoff, Kirchhoff and Cauchy stress spaces respectively:

$$\dot{\mathbf{S}} = {}^{mat}\mathbf{C}^{el} : \left(\dot{\mathbf{E}} - \dot{\mathbf{E}}^{th} - xx \cdot \dot{\mathbf{E}}^{pl}\right) + {}^{mat}\mathbf{C}^{vis} : \left[\ddot{\mathbf{E}} - (1-xx) \cdot \ddot{\mathbf{E}}^{pl}\right], \tag{10}$$

$$\mathcal{L}_P(\mathbf{P}) = \mathbf{F} \cdot \dot{\mathbf{S}} = \mathbf{F} \cdot \left\{ {}^{mat}\mathbf{C}^{el} : \left(\dot{\mathbf{E}} - \dot{\mathbf{E}}^{th} - xx \cdot \dot{\mathbf{E}}^{pl}\right) + {}^{mat}\mathbf{C}^{vis} : \left[\ddot{\mathbf{E}} - (1-xx) \cdot \ddot{\mathbf{E}}^{pl}\right]\right\}, \tag{11}$$

$$\mathcal{L}_O(\boldsymbol{\tau}) = \mathbf{F} \cdot \dot{\mathbf{S}} \cdot \mathbf{F}^T = J \cdot {}^{spat}\mathbf{C}^{el} : \left(\mathbf{d} - \mathbf{d}^{th} - xx \cdot \mathbf{d}^{pl}\right) + J \cdot {}^{spat}\mathbf{C}^{vis} : \left[\mathcal{L}_e(\mathbf{d}) - (1-xx) \cdot \mathcal{L}_e(\mathbf{d}^{pl})\right], \tag{12}$$

$$\mathcal{L}_T(\boldsymbol{\sigma}) = J^{-1} \cdot \mathbf{F} \cdot \dot{\mathbf{S}} \cdot \mathbf{F}^T = {}^{spat}\mathbf{C}^{el} : \left(\mathbf{d} - \mathbf{d}^{th} - xx \cdot \mathbf{d}^{pl}\right) + {}^{spat}\mathbf{C}^{vis} : \left[\mathcal{L}_e(\mathbf{d}) - (1-xx) \cdot \mathcal{L}_e(\mathbf{d}^{pl})\right], \tag{13}$$

where

$$ {}^{mat}\mathbf{C}^{el} = 2 \cdot G \cdot \mathbf{I} + \lambda^{el} \cdot \mathbf{1} \otimes \mathbf{1}, \quad {}^{mat}\mathbf{C}^{vis} = 2 \cdot G^{vis} \cdot \mathbf{I} + \lambda^{vis} \cdot \mathbf{1} \otimes \mathbf{1}, \tag{14}$$

$$G = \frac{E}{2 \cdot (1+\nu)}, \quad \lambda^{el} = \frac{\nu \cdot E}{(1+\nu) \cdot (1-2 \cdot \nu)}, \quad G^{vis} = \frac{E^{vis}}{2 \cdot (1+\nu^{vis})}, \quad \lambda^{vis} = \frac{\nu^{vis} \cdot E^{vis}}{(1+\nu^{vis}) \cdot (1-2 \cdot \nu^{vis})}, \tag{15}$$

$$ {}^{spat}\mathbf{C}^{el}_{ijkl} = J^{-1} \cdot F_{im} \cdot F_{jn} \cdot F_{ko} \cdot F_{lp} \cdot {}^{mat}\mathbf{C}^{el}_{mnop}, \tag{16}$$

$$ {}^{spat}\mathbf{C}^{vis}_{ijkl} = J^{-1} \cdot F_{im} \cdot F_{jn} \cdot F_{ko} \cdot F_{lp} \cdot {}^{mat}\mathbf{C}^{vis}_{mnop}. \tag{17}$$

In Eqns. (10)-(17) the symbols $\mathbf{S}, \mathbf{P}, \boldsymbol{\tau}, \boldsymbol{\sigma}, \dot{\mathbf{S}}, \mathcal{L}_P(\mathbf{P}), \mathcal{L}_O(\boldsymbol{\tau}), \mathcal{L}_T(\boldsymbol{\sigma})$ denote the 2$^{nd}$ Piola-Kirchhoff stress tensor, the 1$^{st}$ Piola-Kirchhoff stress tensor, the Kirchhoff stress tensor, the Cauchy stress tensor and their objective rates. They are a time derivative of the 2$^{nd}$ Piola-Kirchhoff stress tensor $\dot{\mathbf{S}}$, the Lie derivative of the 1$^{st}$ Piola-Kirchhoff stress tensor $\mathcal{L}_P(\mathbf{P})$, defined in terms of the Lie derivative operator of a mixed spatial-material stress tensor $\mathcal{L}_P(\bullet) = \mathbf{F} \cdot \left[\partial\left(\mathbf{F}^{-1} \cdot (\bullet)\right)/\partial t\right]$, the Oldroyd rate of the Kirchhoff stress $\mathcal{L}_O(\boldsymbol{\tau})$, defined in terms of the Lie derivative operator of a spatial stress tensor $\mathcal{L}_O(\bullet) = \mathbf{F} \cdot \left[\partial\left(\mathbf{F}^{-1} \cdot (\bullet) \cdot \mathbf{F}^{-T}\right)/\partial t\right] \cdot \mathbf{F}^T$ and the Truesdell rate of the Cauchy stress $\mathcal{L}_T(\boldsymbol{\sigma})$, defined in terms of the Truesdell derivative operator of a spatial stress tensor $\mathcal{L}_T(\bullet) = J^{-1} \cdot \mathbf{F} \cdot \left[\partial\left(J \cdot \mathbf{F}^{-1} \cdot (\bullet) \cdot \mathbf{F}^{-T}\right)/\partial t\right] \cdot \mathbf{F}^T$, which actually carries out Lie differentiation, but with rearranged terms of its final form Here the fourth order material elasticity tensor ${}^{mat}\mathbf{C}^{el}$ and the fourth order material viscosity tensor ${}^{mat}\mathbf{C}^{vis}$ are defined in the same way as the fourth order material elasticity tensor of the

St.-Venant Kirchhoff material [26], using two independent material parameters, $E, \nu$ and $E^{vis}, \nu^{vis}$ respectively, which ensure isotropy. The fourth order spatial elasticity tensor and the fourth order spatial viscosity tensor $^{spat}\mathbf{C}^{el}$, $^{spat}\mathbf{C}^{vis}$ are then determined in accordance with Eqns. (16)and(17), where $J = \det(\mathbf{F})$ is the Jacobian of the deformation. Here the variable $xx$ denotes the ratio of ductile and total damage increment [25].

To extend the material model for finite-strain cyclic plasticity, Eqns. (10)-(13) have to be supplemented with sufficient evolution equations. The rate forms of the evolution equation defining the backstress tensor when kinematic hardening takes place are based on the NoKH rule of kinematic hardening [27] and can take any of the following forms:

$$^{S}\dot{\mathbf{X}} = {}^{mat}\mathbf{C}^{cyc} : xx \cdot \dot{\mathbf{E}}^{pl} - \gamma\left(e^{el-pl}\right) \cdot {}^{S}\mathbf{X} \cdot \dot{e}^{el-pl}, \tag{18}$$

$$\mathcal{L}_P\left({}^{P}\mathbf{X}\right) = \mathbf{F} \cdot {}^{S}\dot{\mathbf{X}} = \mathbf{F} \cdot \left\{ {}^{mat}\mathbf{C}^{cyc} : xx \cdot \dot{\mathbf{E}}^{pl} - \gamma\left(e^{el-pl}\right) \cdot {}^{S}\mathbf{X} \cdot \dot{e}^{el-pl} \right\}, \tag{19}$$

$$\mathcal{L}_O\left({}^{\tau}\mathbf{X}\right) = \mathbf{F} \cdot {}^{S}\dot{\mathbf{X}} \cdot \mathbf{F}^T = J \cdot {}^{spat}\mathbf{C}^{cyc} : xx \cdot \mathbf{d}^{pl} - \gamma\left(e^{el-pl}\right) \cdot \mathbf{F} \cdot {}^{S}\mathbf{X} \cdot \mathbf{F}^T \cdot \dot{e}^{el-pl}, \tag{20}$$

$$\mathcal{L}_T\left({}^{\sigma}\mathbf{X}\right) = J^{-1} \cdot \mathbf{F} \cdot {}^{S}\dot{\mathbf{X}} \cdot \mathbf{F}^T = {}^{spat}\mathbf{C}^{cyc} : xx \cdot \mathbf{d}^{pl} - \gamma\left(e^{el-pl}\right) \cdot J^{-1} \cdot \mathbf{F} \cdot {}^{S}\mathbf{X} \cdot \mathbf{F}^T \cdot \dot{e}^{el-pl}, \tag{21}$$

where

$$^{mat}\mathbf{C}^{cyc} = 2 \cdot G^{cyc} \cdot \mathbf{I} + \lambda^{cyc} \cdot \mathbf{1} \otimes \mathbf{1}, \quad G^{cyc} = \frac{E^{cyc}}{2 \cdot (1 + \nu^{cyc})}, \quad \lambda^{cyc} = \frac{\nu^{cyc} \cdot E^{cyc}}{(1 + \nu^{cyc}) \cdot (1 - 2 \cdot \nu^{cyc})}, \tag{22}$$

$$^{spat}\mathbf{C}^{cyc}_{ijkl} = J^{-1} \cdot F_{im} \cdot F_{jn} \cdot F_{ko} \cdot F_{lp} \cdot {}^{mat}\mathbf{C}^{cyc}_{mnop}, \quad {}^{P}\mathbf{X} = \mathbf{F} \cdot {}^{S}\mathbf{X}, \quad {}^{\tau}\mathbf{X} = \mathbf{F} \cdot {}^{S}\mathbf{X} \cdot \mathbf{F}^T, \quad {}^{\sigma}\mathbf{X} = J^{-1} \cdot \mathbf{F} \cdot {}^{S}\mathbf{X} \cdot \mathbf{F}^T, \tag{23}$$

$$\gamma\left(e^{el-pl}\right) = \gamma_\infty - (\gamma_\infty - \gamma_0) e^{\left(-\omega \cdot e^{el-pl}\right)}. \tag{24}$$

Here the 4$^{th}$ order material and spatial cyclic-plasticity tensors $^{mat}\mathbf{C}^{cyc}$, $^{spat}\mathbf{C}^{cyc}$ are defined similarly to the elasticity tensors of the material (Eqns. (14)$^1$,(16)), using two material parameters $\nu^{cyc}, E^{cyc}$. The NoKH rule moreover uses three additional material parameters (Eqn.(24)), $\gamma_0, \gamma_\infty, \omega$, which control the backstress tensor contribution into the kinematic hardening and an extra internal state variable $e^{el-pl}$, expressing the ductile damage, which is the plastic damage on the spring in a one-dimensional rheological model of the material (Eqn. (48)$_2$). It ought to be noted here that the objective rates

$\dot{\mathbf{S}}/{}^S\dot{X}, \mathcal{L}_P(\mathbf{P})/\mathcal{L}_P({}^PX), \mathcal{L}_O(\mathbf{\tau})/\mathcal{L}_O({}^\tau X), \mathcal{L}_T(\mathbf{\sigma})/\mathcal{L}_T({}^\sigma X)$ transform in the same way from one form to another as do the stress tensors $\mathbf{S}/{}^S\mathbf{X}, \mathbf{P}/{}^P\mathbf{X}, \mathbf{\tau}/{}^\tau\mathbf{X}, \mathbf{\sigma}/{}^\sigma\mathbf{X}$, which ensure that the formulation is thermodynamically consistent.

### 2.3 On the objectivity and thermodynamic consistency of the formulation

In nonlinear continuum mechanics, a necessary condition of thermodynamic consistency is achieved by the well-known transformations between various stress measures and strain measures respectively [26, 28]. This kind of frame independent representation of the constitutive relations ensures that the important thermodynamic quantities, like the internal energy accumulated in the element in the body's initial and current configurations remain the same during the whole deformation process. The transformations then define necessary conditions of thermodynamic consistency, which in addition imply the following equations, ensuring the equivalence of the rate of change of internal elastic deformation energy Eqn. (25), internal thermal deformation energy Eqn. (26) and internal plastic deformation energy Eqns.(27) and (28) respectively

$$\dot{\mathbf{E}}^{el} : \mathbf{S} \cdot dV_0 = \frac{\partial\,{}^0\dot{\mathbf{u}}^{el}}{\partial \mathbf{X}} : \mathbf{P} \cdot dV_0 = \mathbf{d}^{el} : \mathbf{\tau} \cdot dV_0 = \mathbf{d}^{el} : \mathbf{\sigma} \cdot dv, \tag{25}$$

$$\dot{\mathbf{E}}^{th} : \mathbf{S} \cdot dV_0 = \frac{\partial\,{}^0\dot{\mathbf{u}}^{th}}{\partial \mathbf{X}} : \mathbf{P} \cdot dV_0 = \mathbf{d}^{th} : \mathbf{\tau} \cdot dV_0 = \mathbf{d}^{th} : \mathbf{\sigma} \cdot dv, \tag{26}$$

$$\dot{\mathbf{E}}^{pl} : \mathbf{S} \cdot dV_0 = \frac{\partial\,{}^0\dot{\mathbf{u}}^{pl}}{\partial \mathbf{X}} : \mathbf{P} \cdot dV_0 = \mathbf{d}^{pl} : \mathbf{\tau} \cdot dV_0 = \mathbf{d}^{pl} : \mathbf{\sigma} \cdot dv, \tag{27}$$

$$\dot{\mathbf{E}}^{pl} : {}^S\mathbf{X} \cdot dV_0 = \frac{\partial\,{}^0\dot{\mathbf{u}}^{pl}}{\partial \mathbf{X}} : {}^P\mathbf{X} \cdot dV_0 = \mathbf{d}^{pl} : {}^\tau\mathbf{X} \cdot dV_0 = \mathbf{d}^{pl} : {}^\sigma\mathbf{X} \cdot dv. \tag{28}$$

These then together ensure the equivalence of the rate of change of the overall internal deformation energy also in case of thermal phenomena [28]. Here $dV_0$ is the volume of an infinitesimal volume element in the initial configuration of the body and $dv = J \cdot dV_0$ is its spatial counterpart.

### 2.4 Mathematical modelling of the plastic flow

A proper formulation of the material model in finite-strain plasticity allows for the description of the plastic flow in terms of various instances of a yield surface and stress measures in both the body's initial and current configurations. Let the instances of the yield surface be defined as $^S\Psi = {}^S\Psi\left[{}^S\sigma_{eq}\left(\mathbf{S}, {}^S\mathbf{X}\right), \mathbf{q}\right]$, $^P\Psi = {}^P\Psi\left[{}^P\sigma_{eq}\left(\mathbf{P}, {}^P\mathbf{X}\right), \mathbf{q}\right]$, $^\tau\Psi = {}^\tau\Psi\left[{}^\tau\sigma_{eq}\left(\boldsymbol{\tau}, {}^\tau\mathbf{X}\right), \mathbf{q}\right]$, $^\sigma\Psi = {}^\sigma\Psi\left[{}^\sigma\sigma_{eq}\left(\boldsymbol{\sigma}, {}^\sigma\mathbf{X}\right), \mathbf{q}\right]$ in terms of the stress measures $\mathbf{S}, {}^S\mathbf{X}, /\mathbf{P}, {}^P\mathbf{X}/\boldsymbol{\tau}, {}^\tau\mathbf{X}/\boldsymbol{\sigma}, {}^\sigma\mathbf{X}$ and a vector of internal variables $\mathbf{q}$. After changing the physical interpretation of the plastic flow and applying push-forward and pull-back operations to the material gradient of the plastic velocity field (Eqn. (8)$_1$) as follows

$$\frac{\partial \dot{\mathbf{u}}^{pl}}{\partial \mathbf{x}} = \frac{\partial {}^0\dot{\mathbf{u}}^{pl}}{\partial \mathbf{X}} \cdot \mathbf{F}^{-1} = \dot{\lambda} \cdot \frac{\partial {}^\sigma\Psi}{\partial \boldsymbol{\sigma}} = -\dot{\lambda} \cdot \frac{\partial {}^\sigma\Psi}{\partial {}^\sigma\mathbf{X}}, \quad \frac{\partial \dot{\mathbf{u}}^{pl}}{\partial \mathbf{x}} = \frac{\partial {}^0\dot{\mathbf{u}}^{pl}}{\partial \mathbf{X}} \cdot \mathbf{F}^{-1} = \dot{\lambda} \cdot \frac{\partial {}^\tau\Psi}{\partial \boldsymbol{\tau}} = -\dot{\lambda} \cdot \frac{\partial {}^\tau\Psi}{\partial {}^\tau\mathbf{X}},$$
$$\mathbf{F}^T \cdot \frac{\partial {}^0\dot{\mathbf{u}}^{pl}}{\partial \mathbf{X}} = \dot{\lambda} \cdot \frac{\partial {}^S\Psi}{\partial \mathbf{S}} = -\dot{\lambda} \cdot \frac{\partial {}^S\Psi}{\partial {}^S\mathbf{X}}. \tag{29}$$

it can be found that the yield surfaces are related by the following formulas

$$\frac{\partial {}^P\Psi}{\partial \mathbf{P}} \cdot \mathbf{F}^{-1} = \frac{\partial {}^\sigma\Psi}{\partial \boldsymbol{\sigma}} = -\frac{\partial {}^\sigma\Psi}{\partial {}^\sigma\mathbf{X}}, \quad \frac{\partial {}^P\Psi}{\partial \mathbf{P}} \cdot \mathbf{F}^{-1} = \frac{\partial {}^\tau\Psi}{\partial \boldsymbol{\tau}} = -\frac{\partial {}^\tau\Psi}{\partial {}^\tau\mathbf{X}}, \quad \mathbf{F}^T \cdot \frac{\partial {}^P\Psi}{\partial \mathbf{P}} = \frac{\partial {}^S\Psi}{\partial \mathbf{S}} = -\frac{\partial {}^S\Psi}{\partial {}^S\mathbf{X}}. \tag{30}$$

Equations (29) and (30), supplemented with appropriate formulas for the deformation gradient (Eqn. (3)) and the strain rate tensors (Eqns. (4)-(9)), then serve as a basis for the first nonlinear continuum mechanical theory of finite deformations of thermoelastoplastic media.

Moreover, it can be verified that the yield surfaces and the equivalent stresses $^S\sigma_{eq} = {}^S\sigma_{eq}\left(\mathbf{S}, {}^S\mathbf{X}\right)$, $^P\sigma_{eq} = {}^P\sigma_{eq}\left(\mathbf{P}, {}^P\mathbf{X}\right), {}^\tau\sigma_{eq} = {}^\tau\sigma_{eq}\left(\boldsymbol{\tau}, {}^\tau\mathbf{X}\right), {}^\sigma\sigma_{eq} = {}^\sigma\sigma_{eq}\left(\boldsymbol{\sigma}, {}^\sigma\mathbf{X}\right)$ contained in them, which also comply with the transformations defined by Eqn. (30), have the following properties;

$$^P\sigma_{eq} = {}^S\sigma_{eq} = {}^\tau\sigma_{eq} = J \cdot {}^\sigma\sigma_{eq}, \tag{31}$$

$$\frac{\partial {}^S\Psi}{\partial \mathbf{S}} : \dot{\mathbf{S}} + \frac{\partial {}^S\Psi}{\partial {}^S\mathbf{X}} : {}^S\dot{\mathbf{X}} = \frac{\partial {}^P\Psi}{\partial \mathbf{P}} : \mathcal{L}_P(\mathbf{P}) + \frac{\partial {}^P\Psi}{\partial {}^P\mathbf{X}} : \mathcal{L}_P\left({}^P\mathbf{X}\right) = \frac{\partial {}^\tau\Psi}{\partial \boldsymbol{\tau}} : \mathcal{L}_O(\boldsymbol{\tau}) + \frac{\partial {}^\tau\Psi}{\partial {}^\tau\mathbf{X}} : \mathcal{L}_O\left({}^\tau\mathbf{X}\right) =$$
$$= J \cdot \frac{\partial {}^\sigma\Psi}{\partial \boldsymbol{\sigma}} : \mathcal{L}_T(\boldsymbol{\sigma}) + J \cdot \frac{\partial {}^\sigma\Psi}{\partial {}^\sigma\mathbf{X}} : \mathcal{L}_T\left({}^\sigma\mathbf{X}\right). \tag{32}$$

Besides, since they define the same admissible stress space and the same plastic flow respectively, one of the yield surfaces $^S\Psi, {}^P\Psi, {}^\tau\Psi, {}^\sigma\Psi$ has to be chosen as a reference yield surface to define the material model. It can be shown that when $^\sigma\Psi$ or $^\tau\Psi$ is chosen as a reference yield surface in the current configuration of the body, the contemporary flow plasticity theories will be recovered. It also ought to be noted that Eqns. (30)$_1$ and (30)$_2$ are constraint equations, making the contemporary additive decomposition-based theories appear as if they had mixed finite-strain-small-strain formulation.

A crucial part in finite-strain material modelling is that the formulation of the plastic flow be thermodynamically consistent. This guarantees that plastic deformations are independent from the description used in the model and the particularities of the model formulation. The thermodynamically consistent form of the plastic flow for finite-strain plasticity using combined kinematic and isotropic hardening can then be expressed as

$$\dot{\mathbf{E}}^{pl} : \left( \dot{\mathbf{S}} - {}^S\dot{\mathbf{X}} \right) \cdot dV_0 = \frac{\partial^0 \dot{\mathbf{u}}^{pl}}{\partial \mathbf{X}} : \left[ \mathcal{L}_P(\mathbf{P}) - \mathcal{L}_P\left({}^P\mathbf{X}\right) \right] \cdot dV_0 = \mathbf{d}^{pl} : \left[ \mathcal{L}_O(\boldsymbol{\tau}) - \mathcal{L}_O\left({}^\tau\mathbf{X}\right) \right] \cdot dV_0 =$$
$$= \mathbf{d}^{pl} : \left[ \mathcal{L}_T(\boldsymbol{\sigma}) - \mathcal{L}_T\left({}^\sigma\mathbf{X}\right) \right] \cdot dv. \quad (33)$$

Eqn. (33) has an equivalent form defined by Eqn.(32), known as the 'normality rule', which defines the rate form of a thermodynamically consistent return mapping. The result is of fundamental importance in computational mechanics as it states how the plastic multiplier is to be calculated during return mapping when the plastic step takes place in a finite-strain elastoplastic analysis using a material model with combined isotropic and kinematic hardening.

In order to clarify the physical meaning of Eqn.(32), one ought to take a look at the stress update process in the 2$^{nd}$ Piola-Kirchhoff principal stress space and the other principal stress spaces respectively (see Figure 2)

**Fig. 2**.*Schematic representation of the thermodynamically consistent return-mapping a: in the 2$^{nd}$ Piola-Kirchhoff principal stress space, b: in the 1$^{st}$ Piola-Kirchhoff, Kirchhoff and Cauchy principal stress spaces*

In the 2$^{nd}$ Piola-Kirchhoff stress space, the point characterizing the stress state is fixed in time (see Figure 2a). The incremental forms for the calculation of the 2$^{nd}$ Piola-Kirchhoof stress tensor $\mathbf{S}$ and the corresponding backstress tensor $^{S}\mathbf{X}$ respectively then take the following forms

$$\Delta t \cdot \dot{\mathbf{S}} \approx {}_{\tilde{n}+1}\mathbf{S} - {}_{\tilde{n}}\mathbf{S}, \qquad \Delta t \cdot {}^{S}\dot{\mathbf{X}} \approx {}_{n+1}^{S}\mathbf{X} - {}_{n}^{S}\mathbf{X}, \tag{34}$$

where the left subscripts denote the time step and

$$_{\tilde{n}+1}\mathbf{S} = {}_{\tilde{n}+1}\mathbf{S}(t + \Delta t), \quad {}_{n+1}^{S}\mathbf{X} = {}_{n+1}^{S}\mathbf{X}(t + \Delta t), \tag{35}$$

$$_{\tilde{n}}\mathbf{S} = {}_{\tilde{n}}\mathbf{S}(t) = {}_{\tilde{n}}\mathbf{S}(t + \Delta t), \quad {}_{n}^{S}\mathbf{X} = {}_{n}^{S}\mathbf{X}(t) = {}_{n}^{S}\mathbf{X}(t + \Delta t). \tag{36}$$

In the 1$^{st}$ Piola-Kirchhoff, Kirchhoff and Cauchy stress spaces, the point characterizing the stress state is not fixed in time, due to the fact that the 1$^{st}$ Piola-Kirchhoff, Kirchhoff and Cauchy stress tensors are related to the 2$^{nd}$ Piola-Kirchhoff stress tensor by the formulas $\mathbf{P} = \mathbf{F} \cdot \mathbf{S}$, $\boldsymbol{\tau} = \mathbf{F} \cdot \mathbf{S} \cdot \mathbf{F}^{T}$, $\boldsymbol{\sigma} = J^{-1} \cdot \mathbf{F} \cdot \mathbf{S} \cdot \mathbf{F}^{T}$. The transformations result in a change in the values of the stress tensors whenever the deformation gradient changes, and this also affects the past histories of the tensors. At the beginning of the update process, the stresses $_{n}\mathbf{P}(t), {}_{n}\boldsymbol{\tau}(t), {}_{n}\boldsymbol{\sigma}(t)$, at time step $n$ and time $t$ are represented by the position vector of point C in the 1$^{st}$ Piola-Kirchhoff, Kirchhoff and Cauchy principal stress spaces (see Figure 2b). By the end of the update process, the same stresses, at time step $n$, but time $t + \Delta t$, are then represented by the position vector of point D and values $_{n}\mathbf{P}(t + \Delta t), {}_{n}\boldsymbol{\tau}(t + \Delta t), {}_{n}\boldsymbol{\sigma}(t + \Delta t)$. Similarly, the position vector depicting the backstress tensors $_{n}^{P}\mathbf{X}(t), {}_{n}^{\tau}\mathbf{X}(t), {}_{n}^{\sigma}\mathbf{X}(t)$ at the beginning of the update process, at time step $n$ and time $t$, moves from point A to point B with values of the backstress tensors $_{n}^{P}\mathbf{X}(t + \Delta t), {}_{n}^{\tau}\mathbf{X}(t + \Delta t), {}_{n}^{\sigma}\mathbf{X}(t + \Delta t)$ at the end of the update process.

In addition to this, the Lie derivative of the 1st Piola-Kirchhoff stress tensor $\mathcal{L}_P(\mathbf{P}) = \dot{\mathbf{P}} - \mathbf{L} \cdot \mathbf{P} = \mathbf{F} \cdot \dot{\mathbf{S}}$ implies the following incremental forms for the calculation of the tensors $\mathbf{P}, {}^P\mathbf{X}$ respectively in the 1st Piola-Kirchhoff stress space

$$\Delta t \cdot \mathcal{L}_P(\mathbf{P}) = {}_{n+1}\mathbf{F} \cdot {}_{n+1}\mathbf{S} - {}_{n+1}\mathbf{F} \cdot {}_n\mathbf{S} = {}_{n+1}\mathbf{P}(t+\Delta t) - {}_n\mathbf{P}(t+\Delta t) \quad \text{and}$$
$$\Delta t \cdot \mathcal{L}_P({}^P\mathbf{X}) = {}_{n+1}^{\;P}\mathbf{X}(t+\Delta t) - {}_n^{\;P}\mathbf{X}(t+\Delta t), \tag{37}$$

where

$$_{n+1}\mathbf{P}(t+\Delta t) = {}_{n+1}\mathbf{F} \cdot {}_{n+1}\mathbf{S}, \quad {}_{n+1}^{\;P}\mathbf{X}(t+\Delta t) = {}_{n+1}\mathbf{F} \cdot {}_{n+1}^{\;S}\mathbf{X}, \quad {}_n\mathbf{P}(t) = {}_n\mathbf{F} \cdot {}_n\mathbf{S}, \quad {}_n^{\;P}\mathbf{X}(t) = {}_n\mathbf{F} \cdot {}_n^{\;S}\mathbf{X},$$
$$\mathbf{L} = \dot{\mathbf{F}} \cdot \mathbf{F}^{-1}, \quad {}_n\mathbf{L} = {}_n\dot{\mathbf{F}} \cdot {}_n\mathbf{F}^{-1}, \tag{38}$$

And, by using the Tayor series expansion ${}_n\mathbf{P}(t+\Delta t)$ and ${}_n^{\;P}\mathbf{X}(t+\Delta t)$, can be rewritten as

$${}_n\mathbf{P}(t+\Delta t) = {}_{n+1}\mathbf{F} \cdot {}_n\mathbf{S} \approx {}_n\mathbf{F} \cdot {}_n\mathbf{S} + \frac{\partial(\mathbf{F} \cdot {}_n\mathbf{S})}{\partial t} \cdot \Delta t = {}_n\mathbf{F} \cdot {}_n\mathbf{S} + {}_n\dot{\mathbf{F}} \cdot {}_n\mathbf{S} \cdot \Delta t = {}_n\mathbf{P}(t) + \Delta t \cdot {}_n\mathbf{L} \cdot {}_n\mathbf{P}(t),$$
$${}_n^{\;P}\mathbf{X}(t+\Delta t) = {}_{n+1}\mathbf{F} \cdot {}_n^{\;S}\mathbf{X} = {}_n^{\;P}\mathbf{X}(t) + \Delta t \cdot {}_n\mathbf{L} \cdot {}_n^{\;P}\mathbf{X}(t). \tag{39}$$

Similar incremental forms can be arrived at for the calculation of the Kirchhoff stress tensor $\boldsymbol{\tau}$ and the corresponding backstress tensor ${}^\tau\mathbf{X}$ in the Kirchhoff stress space, using the Oldroyd rate of the Kirchhoff stress $\mathcal{L}_O(\boldsymbol{\tau}) = \dot{\boldsymbol{\tau}} - \mathbf{L} \cdot \boldsymbol{\tau} - \boldsymbol{\tau} \cdot \mathbf{L}^T = \mathbf{F} \cdot \dot{\mathbf{S}} \cdot \mathbf{F}^T$:

$$\Delta t \cdot \mathcal{L}_O(\boldsymbol{\tau}) = {}_{n+1}\mathbf{F} \cdot {}_{n+1}\mathbf{S} \cdot {}_{n+1}\mathbf{F}^T - {}_{n+1}\mathbf{F} \cdot {}_n\mathbf{S} \cdot {}_{n+1}\mathbf{F}^T = {}_{n+1}\boldsymbol{\tau}(t+\Delta t) - {}_n\boldsymbol{\tau}(t+\Delta t),$$
$$\Delta t \cdot \mathcal{L}_O({}^\tau\mathbf{X}) = {}_{n+1}^{\;\tau}\mathbf{X}(t+\Delta t) - {}_n^{\;\tau}\mathbf{X}(t+\Delta t), \tag{40}$$

where

$${}_{n+1}\boldsymbol{\tau}(t+\Delta t) = {}_{n+1}\mathbf{F} \cdot {}_{n+1}\mathbf{S} \cdot {}_{n+1}\mathbf{F}^T, \quad {}_{n+1}^{\;\tau}\mathbf{X}(t+\Delta t) = {}_{n+1}\mathbf{F} \cdot {}_{n+1}^{\;S}\mathbf{X} \cdot {}_{n+1}\mathbf{F}^T,$$
$${}_n\boldsymbol{\tau}(t) = {}_n\mathbf{F} \cdot {}_n\mathbf{S} \cdot {}_n\mathbf{F}^T, \quad {}_n^{\;\tau}\mathbf{X}(t) = {}_n\mathbf{F} \cdot {}_n^{\;S}\mathbf{X} \cdot {}_n\mathbf{F}^T \tag{41}$$

and

$${}_n\boldsymbol{\tau}(t+\Delta t) = {}_{n+1}\mathbf{F} \cdot {}_n\mathbf{S} \cdot {}_{n+1}\mathbf{F}^T \approx {}_n\mathbf{F} \cdot {}_n\mathbf{S} \cdot {}_n\mathbf{F}^T + \frac{\partial(\mathbf{F} \cdot {}_n\mathbf{S} \cdot \mathbf{F}^T)}{\partial t} \cdot \Delta t =$$
$$= {}_n\boldsymbol{\tau}(t) + \Delta t \cdot {}_n\mathbf{L} \cdot {}_n\boldsymbol{\tau}(t) + \Delta t \cdot {}_n\boldsymbol{\tau}(t) \cdot {}_n\mathbf{L}^T,$$
$${}_n^{\;\tau}\mathbf{X}(t+\Delta t) = {}_n^{\;\tau}\mathbf{X}(t) + \Delta t \cdot {}_n\mathbf{L} \cdot {}_n^{\;\tau}\mathbf{X}(t) + \Delta t \cdot {}_n^{\;\tau}\mathbf{X} \cdot {}_n\mathbf{L}^T. \tag{42}$$

And finally, the incremental forms for the calculation of the Cauchy stress tensor $\boldsymbol{\sigma}$ and the corresponding backstress tensor ${}^{\sigma}\mathbf{X}$ can eventually be arrived at, using the Truesdell rate of the Cauchy stress $\mathcal{L}_T(\boldsymbol{\sigma}) = \dot{\boldsymbol{\sigma}} - \mathbf{L}\cdot\boldsymbol{\sigma} - \boldsymbol{\sigma}\cdot\mathbf{L}^T + tr(\mathbf{L})\cdot\boldsymbol{\sigma} = J^{-1}\cdot\mathbf{F}\cdot\dot{\mathbf{S}}\cdot\mathbf{F}^T$ in the Cauchy stress space as follows:

$$\Delta t \cdot \mathcal{L}_T(\boldsymbol{\sigma}) = {}_{n+1}J^{-1} \cdot {}_{n+1}\mathbf{F} \cdot {}_{n+1}\mathbf{S} \cdot {}_{n+1}\mathbf{F}^T - {}_{n+1}J^{-1} \cdot {}_{n+1}\mathbf{F} \cdot {}_n\mathbf{S} \cdot {}_{n+1}\mathbf{F}^T = {}_{n+1}\boldsymbol{\sigma}(t+\Delta t) - {}_n\boldsymbol{\sigma}(t+\Delta t),$$
$$\Delta t \cdot \mathcal{L}_T({}^{\sigma}\mathbf{X}) = {}_{n+1}^{\sigma}\mathbf{X}(t+\Delta t) - {}_n^{\sigma}\mathbf{X}(t+\Delta t), \tag{43}$$

where

$$_{n+1}\boldsymbol{\sigma}(t+\Delta t) = {}_{n+1}J^{-1} \cdot {}_{n+1}\mathbf{F} \cdot {}_{n+1}\mathbf{S} \cdot {}_{n+1}\mathbf{F}^T, \quad {}_{n+1}^{\sigma}\mathbf{X}(t+\Delta t) = {}_{n+1}J^{-1} \cdot {}_{n+1}\mathbf{F} \cdot {}_{n+1}^{S}\mathbf{X} \cdot {}_{n+1}\mathbf{F}^T,$$
$$_n\boldsymbol{\sigma}(t) = {}_nJ^{-1} \cdot {}_n\mathbf{F} \cdot {}_n\mathbf{S} \cdot {}_n\mathbf{F}^T, \quad {}_n^{\sigma}\mathbf{X}(t) = {}_nJ^{-1} \cdot {}_n\mathbf{F} \cdot {}_n^{S}\mathbf{X} \cdot {}_n\mathbf{F}^T, \tag{44}$$

and

$$_n\boldsymbol{\sigma}(t+\Delta t) = {}_{n+1}J^{-1} \cdot {}_{n+1}\mathbf{F} \cdot {}_n\mathbf{S} \cdot {}_{n+1}\mathbf{F}^T \approx {}_nJ^{-1} \cdot {}_n\mathbf{F} \cdot {}_n\mathbf{S} \cdot {}_n\mathbf{F}^T + \frac{\partial(J^{-1}\cdot\mathbf{F}\cdot{}_n\mathbf{S}\cdot\mathbf{F}^T)}{\partial t}\cdot \Delta t =$$
$$= {}_n\boldsymbol{\sigma}(t) + \Delta t \cdot {}_n\mathbf{L} \cdot {}_n\boldsymbol{\sigma}(t) + \Delta t \cdot {}_n\boldsymbol{\sigma}(t) \cdot {}_n\mathbf{L}^T - \Delta t \cdot tr({}_n\mathbf{L}) \cdot {}_n\boldsymbol{\sigma}(t), \tag{45}$$
$$_n^{\sigma}\mathbf{X}(t+\Delta t) = {}_n^{\sigma}\mathbf{X}(t) + \Delta t \cdot {}_n\mathbf{L} \cdot {}_n^{\sigma}\mathbf{X}(t) + \Delta t \cdot {}_n^{\sigma}\mathbf{X}(t) \cdot {}_n\mathbf{L}^T - \Delta t \cdot tr({}_n\mathbf{L}) \cdot {}_n^{\sigma}\mathbf{X}(t).$$

In the rate form of the thermodynamically consistent return mapping procedure defined by Eqn. (32), the objective rates $\dot{\mathbf{S}} / {}^S\dot{\mathbf{X}}$ represent the limits of the vectors pointing from $C$ to $F$ / $A$ to $E$ in the 2$^{nd}$ Piola-Kirchhoff principal stress space as the time step approaches zero $\Delta t \to 0$ (see Figure 2a), while the objective rates $\mathcal{L}_P(\mathbf{P}), \mathcal{L}_O(\boldsymbol{\tau}), \mathcal{L}_T(\boldsymbol{\sigma}) / \mathcal{L}_P({}^P\mathbf{X}), \mathcal{L}_O({}^{\tau}\mathbf{X}), \mathcal{L}_T({}^{\sigma}\mathbf{X})$ represent the limits of the vectors pointing from $D$ to $F$ / $B$ to $E$ in the 1$^{st}$ Piola-Kirchhoff, Kirchhoff and Cauchy principal stress spaces as $\Delta t \to 0$ (see Figure 2b and Eqns. (37), (40) and (43) respectively). The ordinary time derivatives $\dot{\mathbf{S}}, \dot{\mathbf{P}}, \dot{\boldsymbol{\tau}}, \dot{\boldsymbol{\sigma}} / {}^S\dot{\mathbf{X}}, {}^P\dot{\mathbf{X}}, {}^{\tau}\dot{\mathbf{X}}, {}^{\sigma}\dot{\mathbf{X}}$ are in all cases the limits of the vectors pointing from $C$ to $F$ / $A$ to $E$ as $\Delta t \to 0$ and the difference between the ordinary time derivatives and the objective time derivatives of the corresponding pairs of tensors is the limit of the vectors pointing from $C$ to $D$ / $A$ to $B$ respectively. The stress update processes in the 1$^{st}$ Piola-Kirchhoff, Kirchhoff and Cauchy stress spaces can then be viewed as factored versions of the stress update process in the 2$^{nd}$ Piola-Kirchhoff stress space, where the factors are the deformation gradient and the Jacobian, respectively. The replacement of the objective time derivatives by the ordinary time derivatives in Eqn. (32) would result in a thermodynamically inconsistent stress calculation, dependent on the stress measure used in the mathematical model formulation, because the ordinary time derivatives do not comply with the sufficient conditions of thermodynamic consistency mentioned in the above, i.e. $\dot{\mathbf{P}} \neq \mathbf{F}\cdot\dot{\mathbf{S}}, \dot{\boldsymbol{\tau}} \neq \mathbf{F}\cdot\dot{\mathbf{S}}\cdot\mathbf{F}^T, \dot{\boldsymbol{\sigma}} \neq J^{-1}\cdot\mathbf{F}\cdot\dot{\mathbf{S}}\cdot\mathbf{F}^T$.

This is not the case when the objective rates $\dot{\mathbf{S}}, \mathcal{L}_P(\mathbf{P}), \mathcal{L}_O(\boldsymbol{\tau}), \mathcal{L}_T(\boldsymbol{\sigma})$ are used in the return mapping procedure, as $\mathcal{L}_P(\mathbf{P}) = \mathbf{F} \cdot \dot{\mathbf{S}}$, $\mathcal{L}_O(\boldsymbol{\tau}) = \mathbf{F} \cdot \dot{\mathbf{S}} \cdot \mathbf{F}^T$ and $\mathcal{L}_T(\boldsymbol{\sigma}) = J^{-1} \cdot \mathbf{F} \cdot \dot{\mathbf{S}} \cdot \mathbf{F}^T$. The same applies for the backstress tensors ${}^S\dot{\mathbf{X}}, \mathcal{L}_P({}^P\mathbf{X}), \mathcal{L}_O({}^\tau\mathbf{X}), \mathcal{L}_T({}^\sigma\mathbf{X})$. The origin of the objective time derivatives is well understood by considering that corresponding quantities are comoving with the material and the constitutive relations [37].

## 2.5 The reference yield surface

It has been shown in the above that the reference yield surface governs the material model. As a result, contemporary flow plasticity theories can be generalized and alternative material models may be developed. In this research, the $J_2$ plasticity theory using combined kinematic and isotropic hardening is generalized, employing ${}^P\Psi = {}^P\Psi\left({}^P\sigma_{eq}(\mathbf{P}, {}^P\mathbf{X}), \mathbf{q}\right)$ (Eqn.(46)$_1$) as the reference yield surface to define the material model. It should also be noted here that the ${}^PJ_2(\mathbf{P}, {}^P\mathbf{X}) = (\mathbf{P} - {}^P\mathbf{X}) : (\mathbf{P} - {}^P\mathbf{X})$ invariant in the definition of the equivalent stress ${}^P\sigma_{eq}$ is no longer based on the deviatory parts of the tensors $\mathbf{P}$ and ${}^P\mathbf{X}$. This is due to the fact that the 1$^{st}$ Piola-Kirchhoff stress tensor transforms under a change of the observer, while $\mathbf{P}^+ = \mathbf{Q}_R \cdot \mathbf{P}$, ${}^P\mathbf{X}^+ = \mathbf{Q}_R \cdot {}^P\mathbf{X}$ and ${}^PJ_2(\mathbf{P}, {}^P\mathbf{X})$ is the only invariant which is not affected by the change, i.e. ${}^PJ_2(\mathbf{P}, {}^P\mathbf{X}) = {}^PJ_2(\mathbf{P}^+, {}^P\mathbf{X}^+)$. Here $\mathbf{Q}_R$ is an arbitrary rotating tensor expressing the relative rotation of the coordinate system of an observer with respect to the reference coordinate system. The resulting yield surface is then no longer a cylinder, but a sphere.

$$^P\Psi = {}^P\sigma_{eq} - {}^P\sigma_y \leq 0, \quad \text{where} \quad {}^P\sigma_{eq} = {}^P\sigma_{eq}(\mathbf{P}, {}^P\mathbf{X}) = \sqrt{{}^PJ_2(\mathbf{P}, {}^P\mathbf{X})} = \sqrt{(\mathbf{P} - {}^P\mathbf{X}) : (\mathbf{P} - {}^P\mathbf{X})}, \quad (46)$$

$$^P\sigma_y = F_{UT11} \cdot \sqrt{r^2 - \left[a \cdot e^{pl} - \text{center}\right]^2}, \quad r = \sigma_y + Q, \quad \text{center} = \sqrt{r^2 - \sigma_y^2} \quad \text{and} \quad a = \frac{\text{center} + r}{b}, \quad (47)$$

$$\dot{e}^{el-pl} = \dot{e}^{el-pl}\left(xx \cdot \dot{\mathbf{F}}^{pl}\right) = \sqrt{\left(xx \cdot \dot{\mathbf{F}}^{pl}\right) : \left(xx \cdot \dot{\mathbf{F}}^{pl}\right)} = xx \cdot \dot{\lambda}, \quad e^{el-pl} = \int_0^t \dot{e}^{el-pl} \cdot dt, \quad (48)$$

$$\dot{e}^{pl} = \dot{e}^{pl}\left(\dot{\mathbf{F}}^{pl}\right) = \sqrt{\dot{\mathbf{F}}^{pl} : \dot{\mathbf{F}}^{pl}} = \dot{\lambda}, \quad e^{pl} = \int_0^t \dot{e}^{pl} \cdot dt, \quad (49)$$

$$\mathbf{F}^{pl} = \mathbf{I} + \frac{\partial\,^0\mathbf{u}^{pl}}{\partial \mathbf{X}}, \quad \dot{\mathbf{F}}^{pl} = \frac{\partial\,^0\dot{\mathbf{u}}^{pl}}{\partial \mathbf{X}} = \dot{\lambda} \cdot \frac{\partial\,^P \Psi}{\partial \mathbf{P}}. \tag{50}$$

The actual yield stress $^P\sigma_y$, which is a 1st Piola-Kirchhoff stress measure, determines the radius of the yield surface and is defined by Eqn. (47)$_1$. It is the only nonzero component of the stress tensor $\mathbf{P}_{UT}$ (i.e. $^P\sigma_y = [\mathbf{P}_{UT}]_{11}$) coming from the uniaxial tensile test of the modelled material, where the operator $[(\bullet)]_{11}$ extracts the element in the first row and the first column of a 2nd order tensor $(\bullet)$, written as a 3x3 square matrix. The corresponding deformation gradient and the Jacobian of deformation are denoted as $\mathbf{F}_{UT}, J_{UT}$, where $F_{UT11} = [\mathbf{F}_{UT}]_{11}$ and $J_{UT} = \det(\mathbf{F}_{UT})$. It ought to be noted that the only nonzero element of the corresponding 2nd Piola-Kirchhoff stress tensor $\mathbf{S}_{UT}$, coming from the tensile test of the material, is $[\mathbf{S}_{UT}]_{11} = {}^S\sigma_y = \sqrt{r^2 - [a \cdot e^{pl} - \text{center}]^2}$. The equation defines an arc of a circle employing three material parameters, the constant yield stress of the material $\sigma_y$, the maximum stress $Q$, by which the material can harden and the accumulated strain maximum value $b = e^{pl}_{\max}$, at which the material loses its integrity, i.e. $^S\sigma_y = 0$. The relationship between the stress tensors then can be written as $\mathbf{P}_{UT} = \mathbf{F}_{UT} \cdot \mathbf{S}_{UT}$, where the parameters $\sigma_y, Q$ are 2nd Piola-Kirchhoff stress measures and $e^{pl} \subset \langle 0, b \rangle$. It should also be noted here that the definitions of the accumulated plastic strain rates $\dot{e}^{el-pl}, \dot{e}^{pl}$ (Eqns. (48)$_1$ and (49)$_1$) expressing the rate of change of ductile and total damage, and the equivalent stress $^P\sigma_{eq}$ (Eqn. (46)$_2$) respectively have changed. This resulted from the need to meet the requirements of thermodynamic consistency in both a one-dimensional (1D) stress state and a three-dimensional (3D) stress state, which may occur at a material particle in the analysis. Here $\mathbf{F}^{pl}$ denotes the deformation gradient of pure plastic deformations, the time derivative of which is assumed to be in the form of Eqn. (8)$_1$ and $xx$ is the ratio of ductile and total damage increment [25].

## 2.6 The calculation of the plastic multiplier

The calculation of the plastic multiplier is a crucial step in finite-strain elastoplastic stress analysis, as it determines the values of the stress rate tensor (Eqns. (10)-(13)), the backstress rate tensor (Eqns. (18)-(21)) and the plastic part of the strain rate tensor (Eqns. (6), (9)$_3$) when plastic deformations take place in the analysis.

Moreover, the return mapping procedure has to be thermodynamically consistent, i.e. it has to comply with Eqn. (32). This condition has never been met in finite-strain computational plasticity. The thermodynamically consistent return mapping procedure then utilizes the objective differentiation of the yield surface $^P\Psi$ in the form

$$\frac{\partial ^P\Psi}{\partial \mathbf{P}}:\mathcal{L}_P(\mathbf{P})+\frac{\partial ^P\Psi}{\partial ^P\mathbf{X}}:\mathcal{L}_P(^P\mathbf{X})-[\mathcal{L}_P(\mathbf{P}_{UT})]_{11}=0, \tag{51}$$

where $\mathcal{L}_P(\mathbf{P}), \mathcal{L}_P(^P\mathbf{X})$ are then replaced by the rate forms of the constitutive and evolution equations of the material(Eqns.(11) and (19)) and the third term of Eqn. (51)by the expression $[\mathcal{L}_P(\mathbf{P}_{UT})]_{11} = F_{UT11} \cdot \left\{ \left[ -a \cdot (a \cdot e^{pl} - \text{center}) \right] / \sqrt{r^2 - \left[ a \cdot e^{pl} - \text{center} \right]^2} \right\} \cdot \dot{e}^{pl}$. It ought also be noted that the first two terms of Eqn. (51)may be replaced by any other two corresponding terms of Eqn.(32), because the material model formulation is thermodynamically consistent.

**2.7 The ratio of ductile and total damage increment**

The idea of the ratio of ductile and total damage increment $xx$ was first introduced by Écsi and Élesztős in order to properly account for material damping when the plastic step takes place in an elastoplastic analysis. The ratio allows for the redistribution of the plastic flow between the spring and the damper of the 1D frictional device representing the rheological model of the material [25]. The ratio is determined in the elastic predictor phase during return mapping and its value is then kept constant. Since the return mapping procedure in the presented material model is carried out in the 1st Piola-Kirchhoff stress space, the definition of the ratio hasto be modified as follows

$$xx = \frac{\langle \mathbf{N}:\mathbf{F} \cdot [^{mat}\mathbf{C}^{el}:(\dot{\mathbf{E}}-\dot{\mathbf{E}}^{th})] \rangle}{\langle \mathbf{N}:\mathbf{F} \cdot [^{mat}\mathbf{C}^{el}:(\dot{\mathbf{E}}-\dot{\mathbf{E}}^{th})] \rangle + \langle \mathbf{N}:\mathbf{F} \cdot (^{mat}\mathbf{C}^{vis}:\ddot{\mathbf{E}}) \rangle}, \qquad xx \in \langle 0,1 \rangle, \tag{52}$$

where

$$\frac{\partial ^P\Psi}{\partial \mathbf{P}}=\mathbf{N}, \quad \frac{\partial ^P\Psi}{\partial ^P\mathbf{X}}=-\mathbf{N}, \quad \mathbf{N}=\frac{\mathbf{P}-^P\mathbf{X}}{\sqrt{(\mathbf{P}-^P\mathbf{X}):(\mathbf{P}-^P\mathbf{X})}}=\frac{\mathbf{P}-^P\mathbf{X}}{\|\mathbf{P}-^P\mathbf{X}\|}, \tag{53}$$

and

$$\langle y \rangle = \frac{y + |y|}{2} \geq 0, \tag{54}$$

denotes the McCauly's brackets, which return zero if $y < 0$ and $\mathcal{L}_P(\mathbf{P}) = \mathbf{F} \cdot \dot{\mathbf{S}}$. It should be noted that all terms on the right-hand-side of Eqn. (52)$_1$ are objective rates, so that the value of $xx$ is not affected by a change of the observer.

### 2.8 The heat equation

In order to describe the conservation of heat energy at a particle of the body, we have modified our former heat equation [29]. The heat equation accounts for the elastic heating and the dissipation-induced heating in the material, and can be expressed in the following material form

$$\rho_0 \cdot c_p \cdot {}^0\dot{T} + \dot{\mathbf{S}} : \mathbf{E}^{thp} = -\nabla_0 \bullet \mathbf{Q} + R, \tag{55}$$

where

$$R = 0.8 \cdot \left(\mathbf{S}^{el} : xx \cdot \dot{\mathbf{E}}^{pl} + \mathbf{S}^{vis} : \dot{\mathbf{E}}\right), \tag{56}$$

$$\dot{\mathbf{S}}^{el} = {}^{mat}\mathbf{C}^{el} : \left(\dot{\mathbf{E}} - \dot{\mathbf{E}}^{th} - xx \cdot \dot{\mathbf{E}}^{pl}\right), \quad \dot{\mathbf{S}}^{vis} = {}^{mat}\mathbf{C}^{vis} : \left[\ddot{\mathbf{E}} - (1 - xx) \cdot \ddot{\mathbf{E}}^{pl}\right], \tag{57}$$

$$\mathbf{E}^{thp} = diag\begin{bmatrix} \alpha_X & \alpha_Y & \alpha_Z \end{bmatrix} \cdot {}^0T + \frac{1}{2} diag\begin{bmatrix} \alpha_X^2 & \alpha_Y^2 & \alpha_Z^2 \end{bmatrix} \cdot {}^0T^2. \tag{58}$$

Here the symbols $\rho_0, c_p, {}^0T, \dot{\mathbf{S}}, \mathbf{E}^{thp}, \mathbf{Q}, R$ denote the material density, the specific heat at constant pressure, the absolute temperature field, the rate of change of the 2$^{nd}$ Piola-Kirchhoff stress tensor, (Eqn.(10)), a specific Green thermal strain tensor originating from thermal expansion only and expressing the thermal strain with respect to the absolute zero temperature, the material heat flux vector and the heat generation rate per unit volume. The elastic heating in the equation is defined by the term $\dot{\mathbf{S}} : \mathbf{E}^{thp}$ and the dissipation induced heating by the heat generation rate per unit volume as a product of a ratio of dissipated energy converted into heat and the amount of dissipated energy (Eqn. (56)). The typical ratio of plastic work converted into heat is between $0.9 - 1.0$ [30], although this value might differ greatly for different steels with average values between $0.6 - 1.0$ [31]. The ratio is generally assumed to be constant and independent of plastic deformation and the strain rate. In this paper it is assumed that 80% of the dissipated mechanical energy is converted into heat. Here $\mathbf{S}^{el}$ is the elastic part of the stress tensor and $\mathbf{S}^{vis}$ is its viscous part, with rate forms defined by Eqn. (57).

## 3. Numerical experiment

In our numerical experiment a notched, 2024-T3 aluminium alloy specimen in cyclic tension was studied, using $2\ \text{Hz}$ circular frequency and zero stress ratio $R=0$. The amplitude of the prescribed axial deformation at the moving end of the specimen increased linearly from zero to its maximum $^0u_{x\max} = 0.68\ \text{mm}$ in the run-up stage of the experiment and was then kept constant. In the numerical study 1/4 of the body was modelled, employing 2 planes of symmetry. Convective and radiation heat transfer was considered through all free surfaces, applying $273.15\ \text{K}$ environmental temperature and radiation source temperature, respectively. The body was initially at rest with $273.15\ \text{K}$ initial temperature. In order to take into account that the moving grip of the testing machine is always cooled to reduce the heat transferring from the machine into the specimen, the heat transfer coefficient $h$ under the grip was increased $10^{20}$ times. The analysis was run as a transient-dynamic analysis, using approximately $0.005\ \text{s}$ time step size, $4\ \text{s}$ run-up time and $16\ \text{s}$ calculation end time. Table 1 outlines the material properties of the 2024-T3 aluminium alloy specimen used in the numerical experiment. The geometry of the specimen is the same as the geometry of the specimen used by the Pastor et. al. collective in their experiment employing infrared thermography [32].

In order to assess the value of the axial component of the deformation gradient coming from the tensile test of the material $F_{UT11}$, a one-dimensional (1D) rate form of the constitutive equation (Eqn.(10)) was solved for the unknown component of the derivative of the elastic axial displacement field with respect to the axial material coordinate $\partial\,^0u_x^{el}/\partial X$. After neglecting the internal damping and the thermal strains in the material, the rate form of the constitutive equation of the tested material, describing the specific 1D stress state during uniaxial tensile loading, can be expressed as

**Tab. 1.** *Material properties of the 2024-T3 aluminium alloy specimen*

$$^S\dot{\sigma}_y = \dot{S}_{11} = E \cdot \left[ \frac{\partial^0 \dot{u}_x^{el}}{\partial X} \cdot \left( 1 + \frac{\partial^0 u_x^{el}}{\partial X} + \frac{\partial^0 u_x^{pl}}{\partial X} \right) \right], \tag{59}$$

where $^S\dot{\sigma}_y = \dot{S}_{11}$ is the axial component of the 2$^{nd}$ Piola-Kirchoff stress rate tensor and $E$ is the Young's modulus of the material. Furthermore, considering that the accumulated plastic strain rate (Eqn. (49)$_1$) in uniaxial test conditions is $\dot{e}^{pl} = \partial^0 \dot{u}_x^{pl} / \partial X = \dot{\lambda}$, and that its integral is $e^{pl} = \partial^0 u_x^{pl} / \partial X$ (Eqn. (49)$_2$), one can find $F_{UT11}$ as a function of the accumulated plastic strain $e^{pl}$

$$F_{UT11} = 1 + \frac{\partial^0 u_x^{el}}{\partial X} + \frac{\partial^0 u_x^{pl}}{\partial X} = 1 + \left[ -\left(1 + e^{pl}\right) + \sqrt{\left(1 + e^{pl}\right)^2 + 2 \cdot \frac{^S\sigma_y}{E}} \right] + e^{pl}, \tag{60}$$

where $^S\sigma_y = {}^S\sigma_y\left(e^{pl}\right)$ see also Eqn.(47). By utilizing the Poisson effect and Eqn. (60), the lateral components $F_{UT22} = F_{UT33} = 1 - \nu \cdot \left(\partial^0 u_x^{el} / \partial X + \partial^0 u_x^{pl} / \partial X\right)$ of the deformation gradient coming from the tensile stress of the material, the gradient itself $\mathbf{F}_{UT} = diag\left[F_{UT11}, \quad F_{UT22}, \quad F_{UT33}\right]$ and the actual yield stresses $^P\sigma_y, {}^\sigma\sigma_y$, as 1$^{st}$ Piola-Kirchhoff and Cauchy's stress measures can be determined in terms of the accumulated plastic strain $e^{pl}$. Figure 3 depicts the prescribed axial deformation time-history at the moving end of the specimen and the calculated yield stresses ${}^S\sigma_y, {}^P\sigma_y, {}^\sigma\sigma_y$ of the material used in the numerical experiment as 2nd Piola-Kirchhoff, 1st Piola-Kirchhoff and Cauchy's stress measures.

**Fig. 3.**a: The prescribed deformation time-history at the moving end of the specimen, b: The actual yield stresses of the material in [Pa] as functions of the accumulated plastic strain $e^{pl}$ [-]

.

## 4. Numerical results

Figure 4 shows a few selected results taken from the finite element analysis. These are: a, the absolute temperature distribution, b, the axial/x-directional Cauchy's stress distribution and c, the accumulated plastic strain distribution over the volume of the body at the end of the analysis, d, the axial/x-directional Cauchy's stress time histories at nodes N20 and N17 (see Figure 4b for the location of the nodes), e, the temperature change time histories at nodes N20 and N17 and f, the accumulated plastic strain time histories at nodes N17 and N20 respectively.

**Fig. 4.**  *Selected results: a: Absolute temperature distribution, b: Cauchy's stress in the x direction, c: Accumulated plastic strain, d: Cauchy's stress time-history (t.h.) in the x direction at nodes N20, N17, e: Temperature change t.h. at nodes N20, N17, f: Accumulated plastic strain t.h. at nodes N20, N17*

It ought to be noted that, in spite of the fact that the body is not loaded in compression (Figure 3a), after a while compressive axial stresses start to develop in the material which, after a few completed cycles, oscillate as if the loading of the body was symmetric with a stress ratio $R = -1$ (Figure 4d). As a result, both positive and negative residual stresses are created upon unloading (Figure 4b), while the accumulated plastic strain value is continuously increasing at the critical cross-section of the specimen (Figure 4f) with final accumulated plastic strain distribution in Figure 4c and corresponding absolute temperature distribution in Figure 4a at the end of the analysis. It can be seen from the figure that due to cooling there is almost no change in the temperature under the moving grip, while maximum changes take place in the area of the notch. The changes in the time history of the axial stress(Figure4d) result from ratchetting, which causes progressive plastic flow in the direction of the mean stress as the number of loading cycles increases. The temperature decreases in elastic loading and increases when plastic deformations start to take place in the material, with a maximum temperature increase circa 5 C at approximately 7.5 s, while it decreasesagain until the end of the numerical experiment (Figure4e).

The presented theory is also noteworthy from the material testing point of view, as it shows that contemporary tensile testing of ductile materialsis not sufficient for finite-strain material property determination unless the deformation gradient is alsomeasured during the test. The presented theory thus also servesas a basis for improved material testing.

## 5. Conclusions

In this paper a combined objective thermoelastoplastic material model was developed and its performace demonstrated by the numerical calculation of an notched Al specimen subject to cyclic tension. The model is based on the additive decomposition of the strain rate into elastic, plastic and thermal parts, according to a general reference frame independent framework [24,37], but realized here with the help of displacement decomposition. Due to the objective treatment the presented formulation leads to frame independent thermodynamic quantitites . Therefore, the described phenomenon and the results of the analyses employing the model are no longer affected by the particularities of model formulation. The modified nonlinear continuum theory of finite deformations of thermoelastoplastic media furthermore allows for the generalization of contemporary flow plasticity theories and the development of thermodynamically consistent alternative material models while also serving as a basis for improved material testing of real materials, in presence of thermal phenomena.

**Acknowledgements**

This publication is the result of the project implementation: "Research of friction stir welding (FSW) application as an alternative to melting welding methods" no. 26240220031 supported by the Research & Development Operational Programme and funded by the ERDF. Funding from the VEGA grant 1/0740/16 resources and National Research, Development and Innovation Office – NKFIH 116197 NKFIH 124366, NKFIH 116197 is also greatly appreciated.

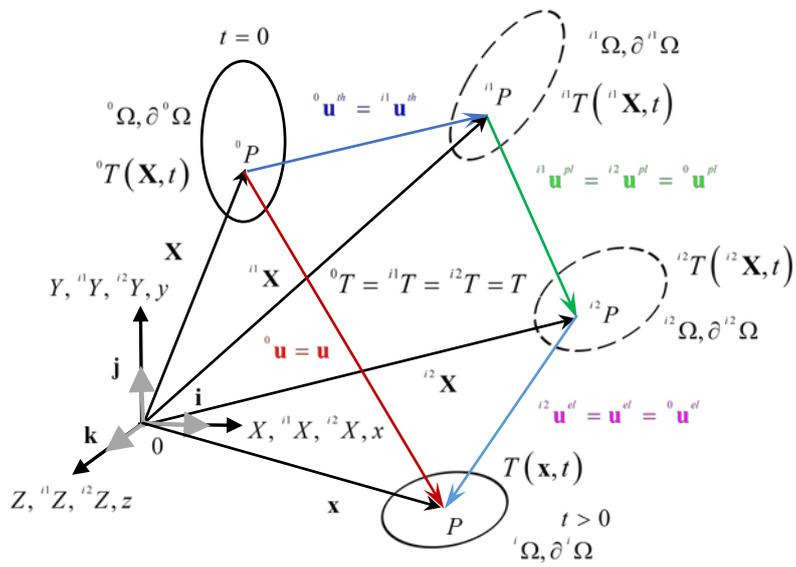

**Fig. 1.** *Multiplicative decomposition of the deformation gradient into an elastic part, a plastic part and a thermal part*

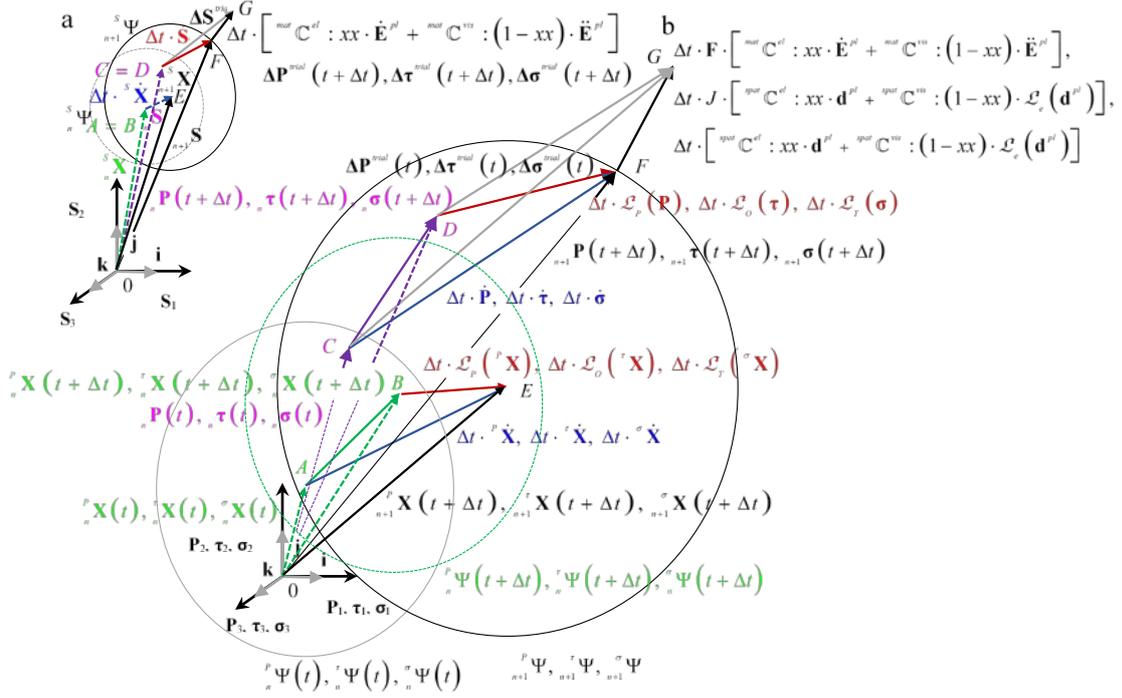

**Fig. 2**. *Schematic representation of the thermodynamically consistent return-mapping a: in the 2nd Piola-Kirchhoff principal stress space, b: in the 1st Piola-Kirchhoff, Kirchhoff and Cauchy principal stress spaces*

**Tab. 1.** *Material properties of the 2024-T3 aluminium alloy specimen*

| | |
|---|---|
| $E$ [Pa] | $7.31 \cdot 10^{10}$ |
| $E^{vis}$ [Pa·s] | $7.31 \cdot 10^{4}$ |
| $E^{cyc}$ [Pa] | $7.31 \cdot 10^{7}$ |
| $\nu = \nu^{vis} = \nu^{cyc}$ [−] | 0.33 |
| $\sigma_y$ [Pa] | $220.0 \cdot 10^{6}$ |
| $Q$ [Pa] | $110.0 \cdot 10^{6}$ |
| $b$ [-] | 1.0 |
| $\gamma_\infty$ [−] | 0.001 |
| $\gamma_0$ [−] | 0.002 |
| $\omega$ [−] | 10.0 |
| $\rho_0$ [kg/m$^3$] | 2770.0 |
| $c$ [J·kg$^{-1}$·°K$^{-1}$] | 876.0 |
| $k_{XX} = k_{YY} = k_{ZZ}$ [W·m$^{-1}$·°K$^{-1}$] | 120.0 |
| $\alpha_X = \alpha_Y = \alpha_Z = \alpha$ [°K$^{-1}$] | $23.4 \cdot 10^{-6}$ |
| $h$ [W·m$^{-2}$·°K$^{-1}$] | 10.0 |
| $\psi$ [-] | 1.0 |
| $\sigma_{EMS}$ [W·m$^{-2}$·°K$^{-4}$] | $11.341 \cdot 10^{-9}$ |

.

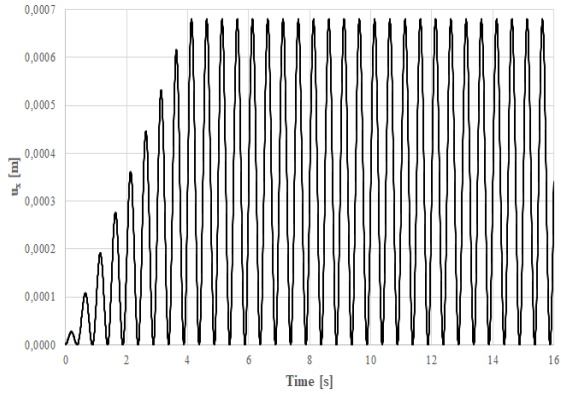 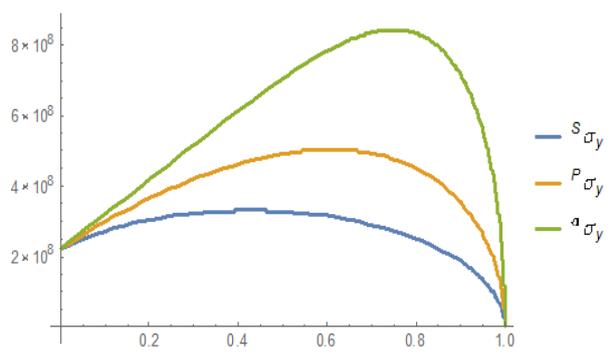

a, b,

**Fig. 3.** *a: The prescribed deformation time-history at the moving end of the specimen, b: The actual yield stresses of the material in [Pa] as functions of the accumulated plastic strain $e^{pl}$ [-]*

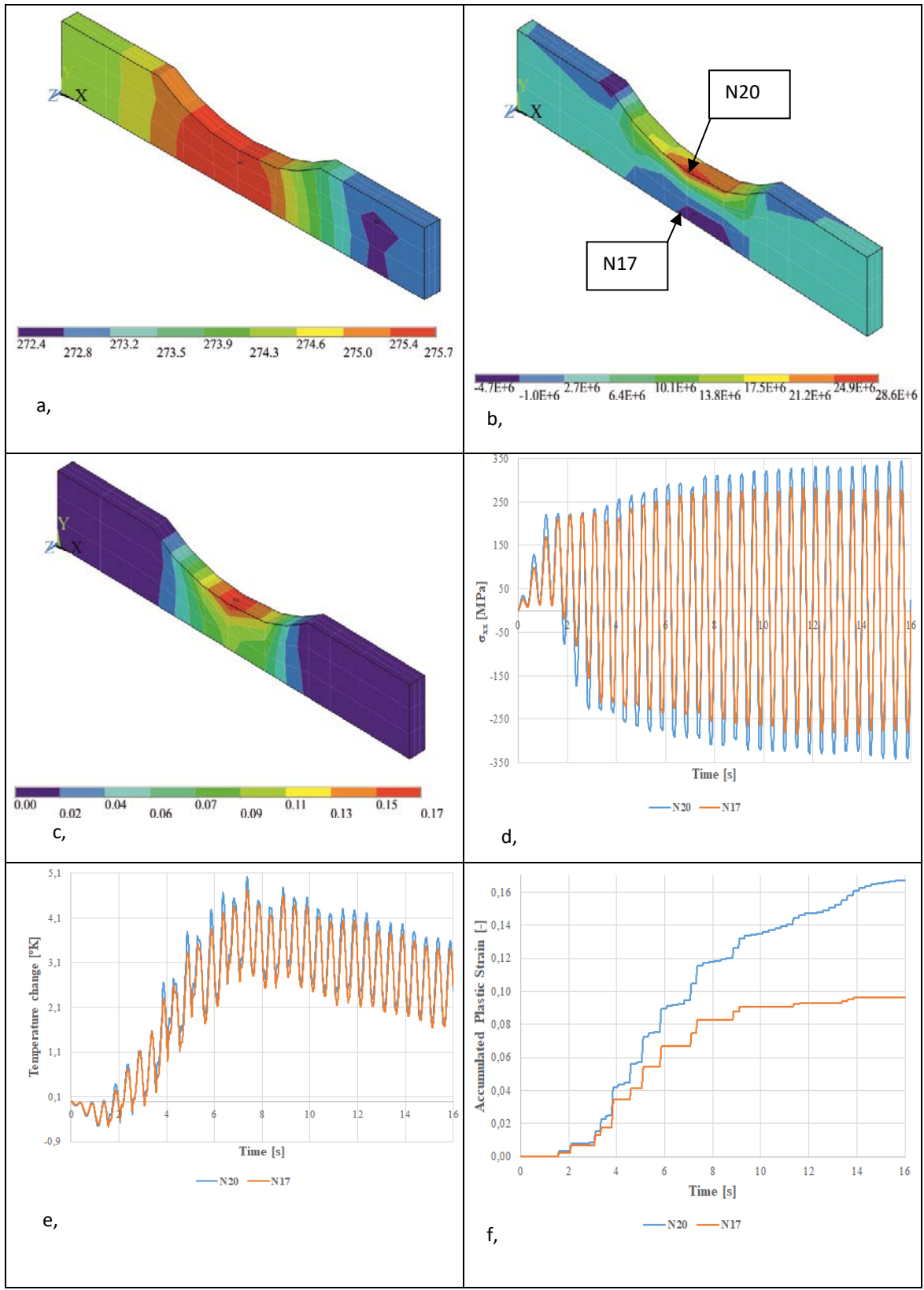

**Fig. 4.** *Selected results: a: Absolute temperature distribution, b: Cauchy's stress in the x direction, c: Accumulated plastic strain, d: Cauchy's stress time-history (t.h.) in the x direction at nodes N20, N17, e: Temperature change t.h. at nodes N20, N17, f: Accumulated plastic strain t.h. at nodes N20, N17*